\documentclass{article}
\pdfoutput=1
\usepackage{graphicx}

\usepackage{subfigure}
\usepackage{amsmath}
\usepackage{mathrsfs} 
\usepackage{amsfonts}
\usepackage{amssymb}%
\usepackage{amsthm}
\usepackage{amscd}
\usepackage{verbatim}
 \usepackage{mathabx}
 \usepackage{dsfont}
 \usepackage[all]{xypic} 
\usepackage{endnotes}
\usepackage{enumerate}
\usepackage{mathtools}
\usepackage{comment}
\usepackage{algorithm}
\usepackage{algcompatible}
\usepackage[noend]{algpseudocode}
\usepackage{tikz}
\usepackage{color}

\newcommand{\mcA}{\mathcal{A}}

\newcommand{\mcC}{\mathcal{C}}

\newcommand{\mcD}{\mathcal{D}}

\theoremstyle{plain}
\newtheorem{theorem}{Theorem}[section]

\theoremstyle{definition}
\newtheorem{defn}[theorem]{Definition}
\theoremstyle{definition}

\theoremstyle{definition}
\newtheorem{example}[theorem]{Example}
\theoremstyle{definition}
\theoremstyle{definition}

\theoremstyle{definition}

\usepackage{hyperref}
\usepackage[affil-it]{authblk}
 
\begin{document}
\title{Combining the Connection Scan Algorithm with Contraction Hierarchies}
\author[1]{Jacob Turner}
\affil[1]{Plotwise, Poortweg 4d, 2612 PA Delft, Netherlands.\\ \href{mailto:email@address}{jacob.turner870@gmail.com}}
\maketitle
\begin{abstract}
Since the first solutions finding minimally weighted routes in weighted digraphs, a plethora of literature has appeared improving the performance of shortest-path queries for use in real-world applications.  In this paper, we detail how an advanced preprocessing technique for routing algorithms (which create objects known as Contraction Hierarchies) may be combined with the connection scan algorithm, an algorithm originally designed to work with public transportation networks using time tables. This provides an improvement over bi-directional Dijkstra or A${}^*$ search on Contraction Hierarchies.
\end{abstract}
\section{Introduction}

One of the first and most fundamental problems in combinatorial optimization is finding the shortest path between two nodes in a network. Solutions to this problem have manifold applications across theoretical and practical disciplines as well as real-world operations and logistics.

One of the most obvious applications is routing problems in highway networks and public transportation, which serves as the main motivation of this paper. While these problems may be solved by generic shortest path algorithms, many new approaches have appeared to cater specifically to the transit routing context. In this context, the graphs to be routed over are known well in advance. Highway networks are very static and this enables the use of preprocessing to speed up shortest path queries. 

In small cases this can be as extreme as computing the shortest path between all pairs. Then a query may be executed by a simple table look-up. However, it is often the case that the size of the network in question is too large to make computing a such a table  feasible. Various different preprocessing techniques have been developed that are both efficient and help achieve speed ups in shortest path computations.

Many of these techniques rely on the idea of grouping edges or nodes according to some measure of importance and exploiting the resulting hierarchical structure or partitioned structure \cite{sanders2005highway,sanders2006engineering,mohring2007partitioning,bauer2010combining}. The algorithm we consider in this paper, called Contraction Hierarchies, is an extreme version of this where each node is in its own hierarchy \cite{geisberger2008contraction}.

Additional considerations can also complicate more generic algorithms. In the public transportation setting, the fact that vehicles cannot move freely but rather according to set departure times may require augmenting the network in some fashion to ensure that generic algorithms return the correct result.

Because of this, new algorithms have been developed with time tables explicitly in mind. Whereas this was a complication initially \cite{bast2009car}, the extra structure given by time tables allows for very efficient algorithms \cite{round-based-public-transit-routing, dibbelt2013intriguingly}. Once of these algorithms, called the Connection Scan Algorithm, will provide the basis for our suggested improvements to Contraction Hierarchies.

\section{Background}

The first suitably general algorithms for finding shortest paths in weighted digraphs began to appear in the early 20th century, with several variants being published around the same time by Bellman, Ford, Moore and others \cite{shimbel1954structure,bellman1958routing,ford1956network, moore1959shortest}. 

The culmination of these algorithms was presented by Edsger Dijkstra \cite{dijkstra1959note} with a running time of $O(|V|^2)$, now bearing his name. This algorithm proved more efficient than its predecessors and several refinements were immediately made to further accelerate its performance. A notable example improves the running time to  $O(|E|+|V|\log{|V|})$ using Fibonacci Heaps, making the fastest known algorithm for this problem (asymptotically speaking). \cite{fredman1987fibonacci}.

Other techniques have been developed to augment Dijkstra's algorithm. One innovation is the use of a heuristic function that tries to estimate the remaining distance to the target at each step of the algorithm \cite{hart1968formal}. Many heuristics have been proposed and work very well in certain situations, e.g. \cite{tung1992multicriteria, mandow2010multiobjective}. The resulting algorithm is called the $A^*$-algorithm. 

A more straightforward technique is the idea of running Dijkstra's algorithm forward from the source and backwards from the target simultaneously. Eventually a common node will be found between the two algorithms and a shortest path may be reconstructed.

In this paper, we consider Contraction Hierarchies. Contraction Hierarchies refer to the result of preprocessing a weighted digraph such that all of its shortest paths satisfy a desirable property. A shortest path computation is done via bi-directional Dijkstra on the preprocessed graph, ignoring those paths which do not satisfy said desirable property. 

The main point of this paper is to demonstrate that Contraction Hierarchies may also be searched with a bi-directional Connection Scan Algorithm rather than Dijkstra's algorithm. This algorithm has several advantages over Dijkstra in run time and memory management. 

In Sections \ref{sec:ch} and \ref{sec:csa}, we give an overview of Contraction Hierarchies and the Connection Span Algorithm, respectively. In Section \ref{sec:combine}, we discuss how to use the Connection Scan Algorithm on Contraction Hierarchies along with a discussion on implementation details.

\section{Contraction Hierarchies} \label{sec:ch}

Contraction Hierarchies may be viewed as a weighted di-graph satisfying a special property. For most weighted di-graphs, some preprocessing must be performed to turn it into a Contraction Hierarchy. It is this preprocessing we first describe.

Let $G=(V, E, w)$ be a weighted di-graph. Every vertex $v\in V$ is given a distinct numeric label. We call these labels \emph{hierarchies}.  We call the map taking vertices $v\in V$ to their respective hierarchies $\ell$.  While these hierarchies may be assigned arbitrarily, several works have focused on different strategies for assigning hierarchies and their advantages, e.g.  \cite{geisberger2012exact,piperno2008search}.

We define the \emph{neighborhood} of a vertex $v$ as all vertices $u\in V$ such that there exists an arc $e\in E$ with $v$ as either the head or tail of $e$. We denote the neighborhood of $v$ by $N(v)$.

The preprocessing stages iterates through the vertices in increasing order of hierarchy. At the a give, we check if the shortest path from $u$ to $w$ has equal distance to the path $u\to v\to w$ (if that path exists). If it does, then an arc is added from $u$ to $w$ with weight equal to the length of the shortest path from $u$ to $w$. We do the same considering the shortest path from $w$ to $u$ as well. We then proceed to the next node with these "shortcuts" added in.

After this arc adding is done for every vertex in $V$, the result is a Contraction Hierarchy. The salient property of Contraction Hierarchies is that for every pair of points $u,v\in V$, there exists a shortest path between them such that there exists no sub-path of the form $w_1\to w_2\to w_3$ where $\ell(w_2)<\ell(w_1),\ell(w_3)$ \cite{geisberger2008contraction}. This is apparent from the preprocessing stage. 

Any weighted di-graph $G=(V, E, w, \ell)$ satisfying this property can thus be called a Contraction Hierarchy. An equivalent statement is that for every $v_1, v_n\in V$, there is a shortest path between them with the form $v_1\to \cdots\to v_k\to\cdots\to v_n$ where $\ell(v_i) <\ell(v_{i+1})$ for all $1\le i\le k-1$ and $\ell(v_j)>\ell(v_{j+1})$ for all $k\le j\le n$. 
We call $v_k$ the \emph{meeting node} in the shortest path. We define a shortest path of this form to be a \emph{shortest path with a meeting node}. It may be the case the meeting node is either the source or target of the path. 

When computing a shortest path using Contraction Hierarchies, bi-directional Dijkstra's algorithm is used with the following modification: In the forward direction, only arcs pointing towards higher hierarchies are considered; in the reverse direction only arc originating from higher hierarchies are considered.

When the two directions converge, they will do so at the meeting node of the route and a shortest path has been found in the Contraction Hierarchy. Once the shortest path in the Contraction Hierarchy is found, the shortest path in the original di-graph must be reconstructed. This requires some bookkeeping on which arcs were added during preprocessing and what shortest path they are a stand-in for.

Since the number of arcs searched by Dijkstra's algorithm is greatly reduced, the preprocessing stage can give significant speed-ups. The preprocessing strikes a nice compromise between no preprocessing and the computation of the pairwise shortest path look-up table in the following sense:

Suppose the preprocessing step also included adding an arc $u\to W$ for $u,w\in N(v)$ if $\ell(u), if \ell(w)<\ell(v)$ and $u\to v\to w$ is a shortest path between them. It is easy to verify that the resulting di-graph would be complete (or a disjoint union of complete digraphs) with the weight of an arc representing the shortest distance from head to tail. This is equivalent to computing the look-up table of all pairwise shortest distances.

\section{The Connection Scan Algorithm}\label{sec:csa}

\begin{figure}
\centering
\hrule
\vspace{.1cm}
\begin{algorithmic}
\State{$T(d)\gets\Delta_d$ for $d\in\mcD$}
\For {$c\in\mcC,$ $time_d(c) \ge \min(\Delta_d)}$
\If{$time_d(c)\ge T(stn_d(c))$}
\If{$time_a(c) < T(stn_a(c))$}
\State{$T(stn_a(c))\gets time_a(c)$}
\EndIf
\EndIf
\If{$time_d(c)\ge\max\{T(s): s\in\mcA\}$}
\State{Break}
\EndIf
\EndFor
\end{algorithmic}
\hrule
\caption{The Connection Scan Algorithm.}
\label{fig:csa}
\end{figure}

The Connection Scan Algorithm (henceforth, CSA) was developed in a different context than Contraction Hierarchies; the aim was to perform shortest path routing for time-table based transit that is common in trains, trams, and other modes of public transit. While modifications can be made to Dijkstra's algorithm can be made to solve this problem (c.f. \cite{pallottino1998shortest}), CSA was developed from the ground up with this situation in mind.

The fundamental object used in CSA is called a \emph{connection}. This is simply an arc in the underlying transportation network along with an associated departure time and arrival time. If we give an station id to each node, then a connection may be represented as four-tuple: departure station id, arrival station id, departure time, arrival time.

CSA assumes that all connections are sorted from earliest departure time to latest departure time, and ties broken arbitrarily. Every connection is then visited sequentially. At each iteration, it is checked if the connection may extend any of set of currently discovered routes. This is done until all the earliest arrival for each arrival station is found or all connections have been scanned.

More formally, suppose $\mcD$ is a set of departure stops, $\mcA$ a set of arrival stops. Let $\mcC$ be the set of connections, ordered as previously mentioned. We also give a set of departure times $\Delta_d$ for $d\in\mcD$ as input. We call a node $n$ \emph{discovered} if a valid route from some $d\in\mcD$ to $n$ has been found. For every node, let $T(s)$ the earliest known arrival time to $s$. This quantity will be updated at every iteration. If $s$ has not yet been discovered, $T(s):=\infty$.

For a connection $c$, let $stn_d(c)$, $stn_a(c)$, $time_d(c)$, and $time_a(c)$ denote the departure station (or tail), arrival station (or head), departure time, and arrival time of $c$, respectively. Then the pseudo-code for CSA is in Figure \ref{fig:csa}.

The reason that this algorithm gives the correct result is because $\mcC$ is ordered by increasing departure time and a connection $c$ cannot be added to a route with final connection $f$ unless $time_a(f)\le time_d(c)$. In the same vein, there is no need to look at connections departing before the earliest departure time, i.e. $\min(\Delta_d)$.

We can state this property more abstractly allowing us to generalize CSA to other situations. We define the \emph{CSA Property} below. For any graph satisfying the CSA Property, the CSA algorithm may be applied and the proof of correctness directly mirrors the original in \cite{dibbelt2013intriguingly}.

We also need a notion of an \emph{objective function} $\omega$ that assigns every path in a di-graph $G=(V,E)$ a score. We call an objective function \emph{monotonic}
if for two paths $p_1$ and $p_2$, where $p_1$ is a sub-path of $p_2$, $\omega(p_1)\le\omega(p_2)$.

\begin{defn}[CSA Property]\label{def:csa_prop}
Given a weighted graph $G=(V,E)$ (directed or not) and a monotonic objective function $\omega$, we say that it satisfies the \emph{CSA Property} if there is a partial ordering on $\prec$ on $E$ such that: For any $u, w\in V$, there exists a  path  $u\to v_1\cdots\to v_n \to w$ minimizing $\omega$ such that $(u,v_1)\preccurlyeq (v_1,v_2)\preccurlyeq\cdots\preccurlyeq (v_n,w)$.
\end{defn}

The generalized form of CSA is given in Figure \ref{fig:generalized_csa}. As before, let $\mcD$ be the set of departure nodes, $\mcA$ the set of arrival nodes.The  $\Delta_d$ are weights input by the user. We assume that $\mcC$ is ordered increasingly with respect to $\prec$ (with ties between equivalent and incomparable arcs broken arbitrarily). The main difference is that now for $s\in V$, $T(s)$ is defined to be the smallest arc $e\in E$ (with respect to $\prec$) such that $e$ is the final arc in a shortest path from $\mcD$ to $s$. If $s$ has not yet been discovered, we define $T(s):=\infty$. Also, let $h(e)$ be the head (departure) of $e\in E$ and $t(e)$ the tail (arrival).

In the original version of CSA, the partial ordering is defined by $c_1\prec c_2$ if $time_d(c_1)< time_d(c_2)$. The objective function $\omega$ simply assigns a path the arrival time of its final arc. Also, we note that the map $T$ as defined for the generalized CSA provides the extra bookkeeping that allows us to reconstruct shortest paths. 

\begin{figure}
\centering
\hrule
\vspace{.1cm}
\begin{algorithmic}
\State{$T(d)\gets\Delta_d$ for $d\in\mcD$}
\For {$e\in E$}
\If{$T(h(e))\preccurlyeq e$}
\If{$w(e) < w(T(t(e)))$}
\State{$T(t(e))\gets e$}
\EndIf
\EndIf
\If{$T(s)\preccurlyeq e$, $\forall s\in \mcA$}
\State{Break}
\EndIf
\EndFor
\end{algorithmic}
\hrule
\caption{The Generalized Connection Scan Algorithm.}
\label{fig:generalized_csa}
\end{figure}

Having seen CSA, bi-directional CSA is a simple improvement. To perform CSA in the backwards direction, one starts at the opposite end of $\mcC$, searching arcs in decreasing order. We maintain a function $T'(s)$ which is the largest arc (with respect to $\prec$) such that $e$ is the initial arc of a shortest path from $s$ to $\mcA$.

In the uni-directional version, we call a node $s\in V$ \emph{closed} if for all $e\in E$ such that $t(e)= s$, $T(s) \preccurlyeq e$. In other words, we have found a shortest path to $s$. In the bi-directional version, we use the same definition to defined a \emph{forward-closed node} and define a \emph{backward-closed node} analogously: $s$ is backwards-closed if for all $e \in E$ with $h(e)=s$,  $e\preccurlyeq T'(s)$

Bi-directional CSA runs the forward and backward version of CSA in lock-step and and halts when a node $n$ that is both forward- and backward-closed is found. It can then reconstruct a shortest path from $\mcD$ to $n$ and from $n$ to $\mcA$. Concatenation then gives a shortest path from $\mcD$ to $\mcA$.

In fact, we may generalize even further. Rather than searching forwards and backwards with respect to the same order, we may use two different orders. Given two orders $\prec_1, \prec_2$, we define a $(\prec_1,\prec_2)$ bi-directional CSA to be the CSA algorithm that searches forward via the order $\prec_1$, and backwards via the order $\prec_2$. The definition of a forward-closed node and backwards-closed node are easily extended. 

Of course, for general orders $\prec_1,\prec_2$, there is no guarantee that a $(\prec_1, \prec_2)$ bi-directional CSA search will produce any route, let alone a shortest one. For transit, this guarantee is only given if $\prec_1=\prec_2=\prec$ as described above. However, this concept will be exactly what we need in Section \ref{sec:combine}.

\subsection{Computational Trade-offs between CSA and Dijkstra}

We briefly outline the computational advantages of CSA over Dijkstra's algorithm. While the updating of nodes with the best incoming arc is exactly the same as in Dijkstra's algorithm, CSA iterates once over every arc in the graph rather than using any sort of queue, as is typical in implementations of Dijkstra's algorithm. 

A simple \textbf{\texttt{for}} loop is more efficient than the operations used in Priority Queues or Fibonacci Heaps. Modern processors are optimized for looping over chunks of memory, not for search and sort algorithms necessary for Dijkstra's algorithm.

Furthermore, since the order in which the arcs are visited is fixed, they can be contiguously allocated in memory for faster looping as a pre-computation step. Dijkstra's algorithm, by contrast, must build an auxiliary data structure during run time to determine the order of vertices to visit. This means that shortest path queries on Contraction Hierarchies improve the running time of Dijkstra's algorithm style queries from $O(|E| +|V|log(|V|))$ to $O(|E|)$ in the worst case.

There is a trade-off however. The data structure built during Dijkstra's algorithm organizes the search space of edges into a tree in a way compatible with the distance from the departure node. In fact, it iteratively approximates the universal covering tree and the distance to every node not a leaf is known. The current leafs to be settled are precisely the nodes in the data structure.

The advantage of all of this is that one does not need to visit every edge or vertex of the graph to know when the shortest route to the arrival vertex has been found and the algorithm can be terminated early. 

In the worst case, of course, Dijkstra's algorithm will be forced to visit edge and vertex. However, early termination means that real world examples need not be usually worst case. CSA on the other hand uses a precomputed tree, which is in fact a linear order. Since this tree has no non-trivial branching, it only allows for a modest reduction in the search space. Observe that every connection departing after the input departure time and before the earliest possible arrival time must be considered.

\section{Applying Connection Scan Algorithm to Contraction Hierarchies}\label{sec:combine}

When considering Definition \ref{def:csa_prop}, it is natural to wonder what other partial orders may arise naturally that satisfy the CSA property. For arbitrary graph routing, there clearly is no choice of ordering that will work. This highlights how special time-table routing is. However, can a graph be modified via some preprocessing such that a canonical order can be found? Serendipitously, it turns out that the preprocessing in making a Contraction Hierarchy does precisely this. 

Given a Contraction Hierarchy $G=(V,E, w, \ell)$, we define two orders on its arcs as follows: For $e_1, e_2\in E$, $e_1\prec_1 e_2$ if $\ell(h(e_1)) < \ell(h(e_2))$, where once again $h(e), t(e)$ denote the head and tail of $e$, respectively. We define the second order by $e_1\prec_2 e_2$ if $\ell(h(e_1)) > \ell(h(e_2))$, i.e. the inverse order of $\prec_1$.

Now suppose that $v_1\to\cdots\to v_n$ is a shortest path with a meeting node from $v_1$ to $v_n$ in $G$. Let $v_k$ be the meeting node of the path. We note that $(v_1, v_2)\prec_1 \cdots \prec_1(v_{k-1}, v_k)$ and $(v_k, v_{k+1})\prec_2 \cdots \prec_2 (v_{n-1}, v_n)$. While routing in Contraction Hierarchies have used a modified bi-directional Dijkstra's algorithm, we see that we may instead use a modified bi-directional CSA algorithm.

Indeed, we search a Contraction Hierarchy via a $(\prec_1,\prec_2)$ bi-directional CSA search, where the forward search only considers only considers arcs $e$ with $\ell(h(e))<\ell(t(e))$ and the backwards search only consider arcs $e$ with $\ell(h(e)) > \ell(t(e))$. We initially define $T(d)=\ell(h(d))$ for $d\in\mcD$ (for the forward search) and $T'(d)=\ell(t(d))$ for $d\in\mcA$ (for the backwards search). This is a slight abuse of the definitions $T$, $T'$, but in this context the meaning is clear. 

Otherwise, $T$ and $T'$ are as they were defined for the Generalized CSA algorithm above, essentially constructing linked lists that allow us to reconstruct the used arcs after the main loop terminates.

We know we are guaranteed to find a node that is both forward- and backwards- closed, namely the meeting node. From there we may reconstruct the shortest route with a meeting node, and from there, a shortest route in the original di-graph that the Contraction Hierarchy was built from.

\subsection{Implementation Details}

Above, we laid out the general idea for using CSA to query Contraction Hierarchies. However, we omitted some crucial details that are needed for making the overall algorithm work.

The most important issue we glossed over is how to sort the arcs given the orders $\prec_1$ and $\prec_2$ to ensure that the resulting output is indeed a shortest path. For an arbitrary pair of orders, the na\"ive solution of maintaining two copies of all of the arcs, each sorted appropriately, may be the only viable option. However, this is not desirable and we can do better for the choice of $\prec_1$, $\prec_2$ that we have made.

The first thing to notice is that the forward and backwards version of CSA searches disjoint subsets of arcs. So we maintain two arrays: The first, $E_u$, are the arcs $e$ such that $\ell(h(e))<\ell(t(e))$. The second array $E_d$, are those arcs with $\ell(h(e))>\ell(t(e))$. The first array is sorted increasingly with respect to $\prec_1$ and the second is ordered increasingly with respect to $\prec_2$.

During the preprocessing phase that builds the Contraction Hierarchy, we build the ordered arrays $E_u$ and $E_d$ as follows. Firstly, we call arcs in $E_u$ \emph{upward arcs}; we call the arcs in $E_d$ \emph{downward arcs}. 

For each $v\in V$, we insert all upwards arcs whose tail is $v$ to $E_u$, inserting it in the proper place so as to maintain the order $\prec_1$. For all downward arcs with head at $v$, we do the same for $E_d$ with respect to inverse order of $\prec_2$ (since the backwards search looks in decreasing order). We then insert the added arcs between neighbors of $v$ appropriately as well.  This adds some small increase in computational time to the preprocessing stage. 

As mentioned before, $E_u$ and $E_d$ should be stored contiguously in memory. Furthermore, these arcs should ideally be read-only. The main reason for this is to have thread-safe sharing of $E_u$ and $E_d$ in software services running many copies of the algorithm simultaneously. Given that the networks being routed over are frequently large, one does not want a separate copy of $E_u$ and $E_d$ for each algorithm instance.

We can now present the pseudo-code that performs one to one shortest path query (i.e. $|\mcD|=|\mcA|=1$) over a Contraction Hierarchy with $E_u$ and $E_d$ already computed.  It is in Figure \ref{fig:final_algo}. We note that as in the original CSA, there is no need to scan arcs whose head and tail are in hierarchies lower than both the departure and arrival node.Given an array $A$ let $A[-1]$ denote the last element in $A$. We use $\texttt{NULL}$ as a stand-in for an invalid value.
\begin{figure}[t]
\centering
\hrule
\vspace{.1cm}
\begin{algorithmic}
\State{$T(d)\gets \ell(h(d))$ for $d\in\mcD$}
\State{$T(a)\gets \ell(h(a))$ for $a\in\mcA$}
\State{$u \gets \min\{e \in E_u \;:\;\ell(t(e))\ge\min\{\ell(d):d\in\mcD\}\}$}
\State{$d \gets \max\{e \in E_u \;:\;\ell(h(e))\le\max\{\ell(d):d\in\mcD\}\}$}
\While {$(u \ne \texttt{NULL})$ or $(d \ne \texttt{NULL})$}
\If{$u\ne \texttt{NULL}$ and $T(h(u))\preccurlyeq_1 u$}
\If{$w(u)<w(T(t(u)))$}
\State{$T(t(u))\gets u$}
\EndIf
\EndIf
\If{$d\ne \texttt{NULL}$ and $T'(t(d))\preccurlyeq_2 d$}
\If{$w(d)<w(T(h(d)))$}
\State{$T(d))\gets d$}
\EndIf
\EndIf
\If{$\ell(t(d)) < \ell(h(u))$}
\State{Break}
\EndIf
\If{$u\ne E_u[-1]$}
\State{$u \gets$ next element in $E_u$}
\Else
\State{$u\gets\texttt{NULL}$}
\EndIf
\If{$u\ne E_d[-1]$}
\State{$d \gets$ next element in $E_d$}
\Else
\State{$d\gets\texttt{NULL}$}
\EndIf
\EndWhile
\end{algorithmic}
\hrule
\caption{The Modified CSA algorithm for querying Contraction Hierarchies.}
\label{fig:final_algo}
\end{figure}

Another interesting case to consider finding the shortest route for each pair of  multiple given departure and  arrival nodes. One could multi-thread or parallelize the above algorithm for each pair, however, this is may not be feasible if $|\mcD|$ and $|\mcA|$ grow large. 

We can solve this problem by augmenting the linked lists $T$, $T'$ such that they don't just contain one arc into each reached vertex, but a rather a list arc, departure node pairs for $T$ and an arc, arrival node pairs for $T'$. That is, $T$ associates to each vertex $v$ a dictionary-like object that assigns to each $d\in\mcD$ the last arc $e$ in the currently shortest known path from $d\to v$. Similarly for $T'$. 

When a new arc $e$ is scanned, we check for which $d\in\mcD$ we have already found a path $p_{d,h(e)}: d\to h(e)$. We then see if $p_{d,h(e)} + e$ is a shorter path from $d\to t(e)$ than the currently known shortest path. If so, we update $T$ accordingly. Similarly for the search in the reverse direction.

\subsection{Early Termination}

As was mentioned at the end of Section \ref{sec:csa}, a downside of CSA is that it is difficult to terminate the algorithm early without visiting many of the edges, in contrast to Dijkstra's algorithm.  

In the original CSA algorithm, we could terminate with we came to connections which departed after the best known earliest arrival time of the arrival node. However, the bi-directional CSA above has no natural analog of this condition.

One may be tempted to end the search when the first meeting node is found. However, meeting nodes need not be unique and this approach will not necessarily give the correct answer. This is in contrast to bi-directional Dijkstra's algorithm where termination can occur when the first meeting node is found. Consider the following example.

\begin{example}\label{ex:scan_all}
Consider the following Contraction Hierarchy:
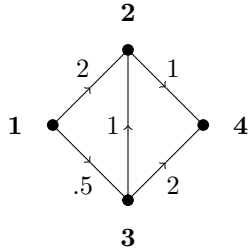
\vspace{5pt}
\begin{center}
\begin{tikzpicture}
\draw[fill=black] (0,0) circle (2pt);
\draw (-.5,0) node{\textbf{1}};
\draw [->] (0,0) -- (.5,.5);
\draw (.5,.5) -- (1,1);
\draw[fill=black] (1,1) circle (2pt);
\draw (1,1.5) node{\textbf{2}};
\draw [->] (0,0) -- (.5,-.5);
\draw[->] (.5,-.5) -- (1,-1);
\draw[->] (1,-1) -- (1,0);
\draw[->] (1,0) -- (1,1);
\draw[fill=black] (1,-1) circle (2pt);
\draw (1,-1.5) node{\textbf{3}};
\draw[->] (1,1) -- (1.5, .5);
\draw (1.5,.5) -- (2,0);
\draw[->] (1,-1) -- (1.5,-.5);
\draw (1.5,-.5) -- (2,0);
\draw[fill=black] (2,0) circle (2pt);
\draw (2.5,0) node{\textbf{4}};
\draw (.4,.75) node{2};
\draw (.4,-.8) node{.5};
\draw (.8, 0) node{1};
\draw (1.6, .75) node{1};
\draw (1.6, -.8) node {2};
\end{tikzpicture}
\end{center}
If the arcs are searched $(1,2)$, $(3,4)$, $(1,3)$, $(2,4)$, $(3,2)$, the shortest path is actually found only after all edges have been searched and two distinct meeting nodes where found, both of which ended up on the shortest path.
\end{example}

As Example \ref{ex:scan_all} above demonstrates, there is not an obvious condition for knowing when the shortest route has been found given the meeting nodes. There are no guarantees how high the hierarchy of the meeting node will be. As such, CSA style querying will likely perform best when all the inputs have high hierarchies as this is the greatest limiter of the search space.

\section{Conclusion}

In essence, Contraction Hierarchies is a preprocessing technique designed to speed up traditional shortest path queries such as bi-directional Dijkstra's algorithm or the bi-directional A${}^*$ algorithm. 

This works by reducing the number of arcs to be considered by the aforementioned methods. Nevertheless, the traditional algorithms used at the querying stage are general purpose and were not designed to take special properties of Contraction Hierarchies into account.

This paper represents an attempt to exploit the salient features of Contraction Hierarchies directly into the algorithm that performs shortest route querying. As it turns out, the fundamental idea behind the Connection Scan Algorithm closely resembles the defining property of a Contraction Hierarchy. Leveraging this, we were able to modify CSA to formulate a new routing algorithm tailored towards Contraction Hierarchies.

The preprocessing stage in making a Contraction Hierarchy removes the need for the auxiliary data structures used in Dijkstra's algorithm, improving the worst-case run time performance by a factor of $O(n\log(n))$. Additionally, the algorithm uses operations for which modern processors are optimized.

The trade-off is that general purpose preprocessing reduces the pruning of the search space that Dijkstra's algorithm provides, forcing the algorithm to visit potentially many more edges. This could have a significant impact in real-world or average-case run-times. 

Open questions are how finer control on the search space can be achieved for the CSA style Contraction Hierarchy queries. Even if for special kinds of graphs, e.g. those satisfying the triangle inequality, this would be of great interest.

\bibliographystyle{plain}
\bibliography{bibfile}

\def\cdprime{$''$} \def\Dbar{\leavevmode\lower.6ex\hbox to 0pt{\hskip-.23ex
  \accent"16\hss}D} \def\cprime{$'$} \def\cprime{$'$} \def\cprime{$'$}
  \def\cprime{$'$} \def\Dbar{\leavevmode\lower.6ex\hbox to 0pt{\hskip-.23ex
  \accent"16\hss}D} \def\cprime{$'$}
\begin{thebibliography}{10}

\bibitem{bast2009car}
Hannah Bast.
\newblock Car or public transport - two worlds.
\newblock In {\em Efficient Algorithms}, pages 355--367. Springer, 2009.

\bibitem{bauer2010combining}
Reinhard Bauer, Daniel Delling, Peter Sanders, Dennis Schieferdecker, Dominik
  Schultes, and Dorothea Wagner.
\newblock Combining hierarchical and goal-directed speed-up techniques for
  dijkstra's algorithm.
\newblock {\em Journal of Experimental Algorithmics (JEA)}, 15:2--3, 2010.

\bibitem{bellman1958routing}
Richard Bellman.
\newblock On a routing problem.
\newblock {\em Quarterly of applied mathematics}, 16(1):87--90, 1958.

\bibitem{round-based-public-transit-routing}
Daniel Delling, Thomas Pajor, and Renato Werneck.
\newblock Round-based public transit routing.
\newblock Society for Industrial and Applied Mathematics, January 2012.

\bibitem{dibbelt2013intriguingly}
Julian Dibbelt, Thomas Pajor, Ben Strasser, and Dorothea Wagner.
\newblock Intriguingly simple and fast transit routing.
\newblock In {\em International Symposium on Experimental Algorithms}, pages
  43--54. Springer, 2013.

\bibitem{dijkstra1959note}
Edsger~W Dijkstra.
\newblock A note on two problems in connexion with graphs.
\newblock {\em Numerische mathematik}, 1(1):269--271, 1959.

\bibitem{ford1956network}
Lester~R Ford~Jr.
\newblock Network flow theory.
\newblock Technical report, RAND CORP SANTA MONICA CA, 1956.

\bibitem{fredman1987fibonacci}
Michael~L Fredman and Robert~Endre Tarjan.
\newblock Fibonacci heaps and their uses in improved network optimization
  algorithms.
\newblock {\em Journal of the ACM (JACM)}, 34(3):596--615, 1987.

\bibitem{geisberger2008contraction}
Robert Geisberger, Peter Sanders, Dominik Schultes, and Daniel Delling.
\newblock Contraction hierarchies: Faster and simpler hierarchical routing in
  road networks.
\newblock In {\em International Workshop on Experimental and Efficient
  Algorithms}, pages 319--333. Springer, 2008.

\bibitem{geisberger2012exact}
Robert Geisberger, Peter Sanders, Dominik Schultes, and Christian Vetter.
\newblock Exact routing in large road networks using contraction hierarchies.
\newblock {\em Transportation Science}, 46(3):388--404, 2012.

\bibitem{hart1968formal}
Peter~E Hart, Nils~J Nilsson, and Bertram Raphael.
\newblock A formal basis for the heuristic determination of minimum cost paths.
\newblock {\em IEEE transactions on Systems Science and Cybernetics},
  4(2):100--107, 1968.

\bibitem{mandow2010multiobjective}
Lawrence Mandow and Jos{\'e} Luis~P{\'e}rez De~La~Cruz.
\newblock Multiobjective a* search with consistent heuristics.
\newblock {\em Journal of the ACM (JACM)}, 57(5):27, 2010.

\bibitem{mohring2007partitioning}
Rolf~H M{\"o}hring, Heiko Schilling, Birk Sch{\"u}tz, Dorothea Wagner, and
  Thomas Willhalm.
\newblock Partitioning graphs to speedup dijkstra's algorithm.
\newblock {\em Journal of Experimental Algorithmics (JEA)}, 11:2--8, 2007.

\bibitem{moore1959shortest}
Edward~F Moore.
\newblock The shortest path through a maze.
\newblock In {\em Proc. Int. Symp. Switching Theory, 1959}, pages 285--292,
  1959.

\bibitem{pallottino1998shortest}
Stefano Pallottino and Maria~Grazia Scutella.
\newblock Shortest path algorithms in transportation models: classical and
  innovative aspects.
\newblock In {\em Equilibrium and advanced transportation modelling}, pages
  245--281. Springer, 1998.

\bibitem{piperno2008search}
Adolfo Piperno.
\newblock Search space contraction in canonical labeling of graphs.
\newblock {\em arXiv preprint arXiv:0804.4881}, 2008.

\bibitem{sanders2005highway}
Peter Sanders and Dominik Schultes.
\newblock Highway hierarchies hasten exact shortest path queries.
\newblock In {\em European Symposium on Algorithms}, pages 568--579. Springer,
  2005.

\bibitem{sanders2006engineering}
Peter Sanders and Dominik Schultes.
\newblock Engineering highway hierarchies.
\newblock In {\em European Symposium on Algorithms}, pages 804--816. Springer,
  2006.

\bibitem{shimbel1954structure}
Alfonso Shimbel.
\newblock Structure in communication nets.
\newblock In {\em Proceedings of the symposium on information networks},
  volume~4. Polytechnic Institute of Brooklyn Nueva York, 1954.

\bibitem{tung1992multicriteria}
Chi~Tung Tung and Kim~Lin Chew.
\newblock A multicriteria pareto-optimal path algorithm.
\newblock {\em European Journal of Operational Research}, 62(2):203--209, 1992.

\end{thebibliography}
\end{document}